\documentstyle[11pt,aaspp4]{article}

\slugcomment{\it Astron.\ J.\ (1999) in press}

\begin{document}

\title{CLASS B1152+199 and B1359+154: Two New Gravitational Lens Systems
Discovered in the Cosmic Lens All-Sky Survey}

\author{S.T. Myers, D. Rusin}
\affil{University of Pennsylvania, Dept. of Physics \& Astronomy,
209 S. 33rd St., Philadelphia, PA, 19104-6396}
\author{C.D. Fassnacht\altaffilmark{1}, R.D. Blandford, T.J. Pearson, 
A.C.S. Readhead}
\affil{Palomar Observatory, 105-24, California Institute of Technology, 
Pasadena CA 91125}
\author{N. Jackson, I.W.A. Browne, D.R. Marlow\altaffilmark{2}, P.N. Wilkinson}
\affil{University of Manchester, NRAL, Jodrell Bank, Macclesfield, 
   Cheshire SK11 9DL, UK}
\author{L.V.E. Koopmans}
\affil{Kapteyn Astronomical Institute, Postbus 800, 9700 AA Groningen, 
The Netherlands}
\and

\author{A.G. de Bruyn\altaffilmark{3}}
\affil{NFRA, Postbus 2, 7990 AA Dwingeloo, The Netherlands}

\altaffiltext{1}{Current address National Radio Astronomy Observatory,
P.O. Box 0, Socorro, NM 87801}
\altaffiltext{2}{Current address University of Pennsylvania, Dept. of Physics
\& Astronomy, 209 S. 33rd St., Philadelphia, PA, 19104-6396}
\altaffiltext{3}{and Kapteyn Astronomical Institute, Postbus 800, 
9700 AA Groningen, The Netherlands}

\begin{abstract}

The third phase of the Cosmic Lens All-Sky Survey (CLASS) has recently been
completed, bringing the total number of sources imaged to over 15000 in
the CLASS and JVAS combined survey.  In the VLA observations carried out in
March and April of 1998, two new candidate lensed systems were discovered:
CLASS B1152+199 and B1359+154.  B1152+199 is a $1\farcs6$ double, with a
background quasar at $z=1.019$ lensed by a foreground galaxy at $z=0.439$.
The relatively flat radio spectra of the lensed images 
($\alpha_{8.46}^{14.94} = -0.32$), combined with a previous ROSAT detection
of the source, make B1152+199 a strong candidate for time delay studies at 
both radio and X-ray wavelengths.  B1359+154 is a quadruply lensed quasar at 
$z=3.235$, with a maximum image separation of $1\farcs7$.  As yet, the 
redshift of the lensing object in this system is undetermined. The steep 
spectral index of the source ($\alpha_{8.46}^{14.94} = -0.9$) suggests that
B1359+154 will not exhibit strong variability, and is therefore unlikely to 
be useful for determining $H_0$ from measured time delays.

\end{abstract}


\section{Introduction} \label{sec:intro}

CLASS, the Cosmic Lens All-Sky Survey, commenced observations in 1994 (Myers
et al.\ 1995), building upon the foundation of JVAS, the Jodrell-VLA
Astrometric Survey (Patnaik et al.\ 1992\markcite{patnaik92}).  In the three
observing ``seasons'' since that time, CLASS has observed over 10000 targets
selected as flat-spectrum between 4.8~GHz and 1.4~GHz, bringing the CLASS
total (including JVAS) to over 15000 sources.  Details of the CLASS survey and
statistical sample will be presented in a forthcoming paper (Myers
et al.\ 1999, in preparation).  In this paper, we present the discovery of two
new lenses from the CLASS-3 observations.

As in the other CLASS papers, we adopt the convention
\begin{equation}
 S \propto \nu^\alpha
\end{equation}
where $\nu$ is the frequency of observation.   Our flat-spectrum sample 
is defined as those sources with $\alpha \geq -0.5$ between 4.85 GHz in the
improved Green Bank survey (GB6; Gregory et al. 1996\markcite{gb6}) and
1.4 GHz in the NRAO-VLA Sky Survey (NVSS; Condon et al.\ 1998\markcite{nvss}).

The CLASS-3 observations mainly concentrated on the region from $0^\circ$ to
$+28^\circ$ declination.  A lower density of fainter sources filled in gaps
in the rest of the CLASS survey area from $+28^\circ$ to $+76^\circ$
declination using the new NVSS selection.  With CLASS-3, nearly all GB6
sources 30 mJy or brighter satisfying the spectral selection have been
observed, with the exception of a small number ($\sim 700$) falling in the
remaining gaps of NVSS.

\section{Observations} \label{sec:obs}


The third phase of the survey (CLASS-3) was carried out with the NRAO Very
Large Array\footnote{The National Radio Astronomy Observatory is operated by
Associated Universities, Inc., under cooperative agreement with the National
Science Foundation.} (VLA) in the A-configuration during March -- May 1998.
The VLA default X-band IFs were used in CLASS-3: $8.4351$~GHz and
$8.4851$~GHz, with an average frequency of $8.46$~GHz.  As in
previous CLASS sessions, 30-second integrations were taken on each source,
with phase calibrators (selected from JVAS) inserted in the schedule after
every 10th CLASS target.  On average, 54 CLASS sources were observed per hour.
A total of 5096 targets were observed in CLASS-3.

The data reduction pipeline was identical to that used in CLASS-1 and CLASS-2
(eg. Myers et al.\ 1995\markcite{myers95}).  Calibration was performed in AIPS,
with mapping automatically carried out in DIFMAP (Shepherd, Pearson \& Taylor
1994\markcite{difmap}).  At the end of the ``automap'' procedure,
model-fitting was performed by using Gaussian components placed at the
positions of the individual small ``clean'' boxes found in the main search
loops.  As the goal of CLASS is to search for the lensing of compact radio
cores, any sources exhibiting multiple compact model-fit components were
identified promptly after sections of the observations had been completed to
allow follow-up in later parts of CLASS-3.

The preliminary automap output from CLASS-3 indicated two excellent lens
candidates: CLASS B1152+199 and B1359+154.  B1152+199 is a $1\farcs6$ double,
with two distinct compact components.  The map of B1359+154 shows five
components, with four compact and one extended in the X-band snapshot.  This
is believed to be a $1\farcs7$ quadruple lens system where lensing galaxy
emission is also seen, making it similar to CLASS B2045+265 
(Fassnacht et al.\ 1998\markcite{2045}).  Follow-up VLA observations of the
two candidates at U band ($14.94$~GHz) were carried out in the April run,
confirming the multiplicity, compactness, and identical component spectra
expected for the lensing hypothesis.

Analysis and model-fitting with point and Gaussian components was done in
DIFMAP.  DIFMAP functions by working on the residual ``difference map''
produced by transforming the residual {\it uv} data after subtraction of the
current model from the original {\it uv} dataset (see Shepherd, Pearson \&
Taylor 1994\markcite{difmap} and Shepherd 1997\markcite{difmap2} for more
details on DIFMAP).  Maps were made using both uniform {\it uv} weighting (to
get the highest resolution possible, at the cost of noise level) and natural
{\it uv} weighting (to get the best signal-to-noise ratio possible, at the
cost of resolution).  Uncertainties in the positions and flux densities were
determined using the square-root of the diagonal elements of the covariance
matrix that DIFMAP computes during the model-fitting procedure.  For the flux
densities, this is very nearly identical to the rms in the naturally-weighted
residual image, and this is what we present in the tables below.
Images were constructed using the residual maps as a noise floor, and
then adding the model components convolved with a Gaussian restoring beam of
width appropriate to the core of the synthesized (``dirty'') beam for the
chosen weighting.

The overall flux density scale used for the X-band observations was based on
$5.18 \pm 0.10$ Jy for J1331+305 (3C286) and $3.22 \pm 0.10$ Jy for J0137+331
(3C48).  This calibration should be good to $\pm 3\%$ or better, as estimated
from the scatter in VLA calibrator measurements over the time period of
CLASS-3 (this information is available from the VLA).  In addition to these
primary calibrators, we used compact steep-spectrum sources J0713+438
($1.20\pm0.02$ Jy) and J1945+709 ($0.43\pm0.01$ Jy).  The gain stability of
the VLA at X band is superb, and the internal consistency between the
correlator counts per Jansky as determined by these independent sources
was better than 2\% for the 5 April 1998 data, for example.  At U band,
however, the flux density scale is more uncertain.  We adopted $3.4\pm0.1$ Jy
for J1331+305 and as a secondary calibrator assumed $0.73 \pm 0.04$ Jy for
J0713+438.  We estimate an overall calibration accuracy of $\pm 3\%$ at 
X band, and $\pm 5\%$ at U band, from the
scatter in VLA calibrator measurements during the several months of the
CLASS-3 observations.  In
the following discussion of uncertainties, and in the error estimates given
in the tables, we do not include these overall scale errors.  However, these
should be kept in mind, especially in the quoted spectral indices between
X band and U band, as the conservative 3\% and 5\% estimated uncertainties
in the X-band and U-band flux densities translate to a spectral index 
uncertainty of $\sigma_\alpha = 0.1$ when propagated.

\subsection{Optical Follow-Up} \label{sec:obs-opt}

Both lens candidates were observed with the Low Resolution Imaging
Spectrograph (LRIS; Oke et al.\ 1995\markcite{lris}) on the W.\ M.\ Keck~II
Telescope on 1998 April 21.  The 1\farcs5 long-slit and 300 gr/mm grating were
used for the observation.  The wavelength coverage was 4024\AA -- 9012\AA,
with a scale on the detector of 2.44\AA/pix.  For each lens candidate the long
slit position angle was chosen to cover the lensed images and the predicted
position of the lensing galaxy.  The data were reduced using standard
IRAF\footnote{IRAF (Image Reduction and Analysis Facility) is distributed by
the National Optical Astronomy Observatories, which are operated by the
Association of Universities for Research in Astronomy under cooperative
agreement with the National Science Foundation.} routines.  The flat-field
frames were constructed from dome flat observations. The sky lines in each
spectrum were used to make corrections for flexure in the spectrograph.
Observations of the Oke standard star Feige 34 (Oke 1990\markcite{okestds})
were used to determine the response of the detector. The spectra were
extracted using the optimal weighting routines in IRAF.

In addition, each candidate was observed with the COSMIC camera on the
Palomar 5-m telescope on 1998 May 31.  A set of images was taken in the Gunn
$g$ and $i$ filters.  The observing conditions were not photometric, with high
cirrus present throughout the evening.  The average seeing was found to be
1\farcs1.  The standard star Feige 67 was observed to set the magnitude scale.
Magnitudes were derived using the IRAF ``apphot'' package.  We estimate
errors in calibration of as much as several tenths of a magnitude given
the conditions.

\section{CLASS B1152+199} \label{sec:1152}

CLASS B1152+199 was selected as GB6 J11553+1939 with a 4.85~GHz flux density
of 76 mJy.  The corresponding 1.4~GHz flux density in the NVSS is 77.4 mJy,
and thus the overall two-point spectral index is $\alpha_{1.4}^{4.85} =
-0.02$.  The target GB6 J11553+1939 also corresponds to MG2 J115518+1940
(Langston et al.\ 1990\markcite{mg2}), which is identified with a mag
$16.9$ optical object in the Palomar Sky Survey E plates
(Brinkmann et al.\ 1997\markcite{brink}), consistent with what we find in our
follow-up. Note that Brinkmann
et al. were observing counterparts to X-ray sources seen in the ROSAT All-Sky
Survey, and located a 69 mJy radio source at 4.85 GHz (VLA D configuration
1992) in this location (Laurent-Muehleisen et al.\ 1997\markcite{laurent}).

Examination of the ROSAT All-Sky Survey catalogue identifies CLASS B1152+199
with RXS J115517.6+193935.  Analysis of the ROSAT data gives a
(0.5--2 keV) flux density of $8 \times 10^{-13}\rm\ erg \,cm^{-2} sec^{-1}$.
ROSAT detection of X-ray emission at this position argues for quasar 
identification.  The bright X-ray emission, combined with the compact 
flat-spectrum radio structure, strongly suggest that this source will be 
variable and therefore a good target for both radio and X-ray time delay
monitoring.

The X-band discovery 30-second snapshot and U-band 5-minute image are shown in
Fig.~\ref{fig:1152}.  Because this is a relatively bright source, we show the
uniformly weighted images, with resolutions of 0\farcs21 (X band) and
0\farcs13 (U band).  We find two components with a relative separation of
$1\farcs56$ and a flux density ratio of 3:1.  Further examination reveals a
third faint radio source in the field.  This object is 23\arcsec\ from the
two compact components, and is probably unrelated (in a lensing sense) to 
the flat spectrum double.  This source is designated as component C in the 
tables.  Positional and flux density data for B1152+199 are listed in
Table~\ref{tab:1152}. 

The combined flux densities of components A and B in the VLA maps are
$69.5$~mJy at $8.46$~GHz, and $58.1$~mJy at $14.94$~GHz.  Thus, the overall
spectral index between the X and U bands is $\alpha_{8.46}^{14.94} = -0.32$,
slightly steeper than the spectral index found at lower frequencies.  Both
components A and B are compact in the uniformly weighted X-band and U-band
images, and the spectral indices of the components are identical within the
uncertainties.

Although the distant component C was extremely faint in the U-band image,
model-fitting was able to solve for a component which converged to the same
position as in the X-band data.  The spectral index of source C between the X
band and U band is significantly steeper than that of components A and B, and
therefore it is unlikely to be another image.  However, it might be the
compact hot-spot in a lobe or an altogether unrelated source.

\subsection{Optical Follow-up of B1152+199} \label{sec:1152-opt}

Spectra of B1152+199 were extracted from two 600-second observations using a
single slit position (PA $= 320\arcdeg$), with lines from both source and lens
seen.  Fig.\,\ref{fig:1152spec1} is dominated
by light from the background source.  Three emission lines from the source are
observed: strong broad \ion{Mg}{2} $\lambda$ 2800\AA\ at 5650\AA, [\ion{O}{2}]
$\lambda$ 3727\AA\ at 7526\AA, and H$\gamma$ at 8760\AA.  These lines
establish the source redshift at $z_s = 1.0189 \pm 0.0004$.  In addition, at
the blue end of the spectrum \ion{Mg}{2} and \ion{Mg}{1} absorption lines
associated with the lens are seen.  A second spectrum was extracted offset
along the slit from the main emission (Fig.\,\ref{fig:1152spec2}), and is
dominated by light from the lensing galaxy.  The features associated with the
lens include the \ion{Mg}{1} absorption seen in Fig.\,\ref{fig:1152spec1} at
4104\AA, [\ion{O}{2}] $\lambda$ 3727 emission at 5368\AA, and \ion{Ca}{2} H
and K absorption at 5660 and 5706\AA.  These lines give a consistent lens
redshift of $z_{\ell} = 0.4386\pm 0.0008$.  Note that the spectrum around the
Ca H and K lines is distorted by the broad \ion{Mg}{2} emission line
from the source.  Several narrow emission lines associated with the
background source are also seen in Fig.\,\ref{fig:1152spec2}, including
\ion{Ne}{5} $\lambda$ 3346\AA\ at 6756\AA, \ion{Ne}{5} $\lambda$ 3427\AA\ at
6918\AA, and the [\ion{O}{2}] and H$\gamma$ observed in Fig.\,\ref{fig:1152spec1}.

The $g$ band COSMIC image of B1152+199 is shown in Fig.\,\ref{fig:1152img}.
There is a bright stellar object ($g = 16.5$ and $i = 16.6$) located at the
radio position.  Presumably, this is the brighter of the two lensed images,
with the weaker image significantly attenuated by dust extinction from the
lens.  Given that the Keck spectrum clearly shows two system redshifts, with
the brighter background source spectrum consistent with that of a $z_s =
1.019$ quasar, it is unlikely that this is simply a chance mis-identification.
Improved ground-based imaging or HST observations will be required to clearly
sort out the optical properties of this system.

\subsection{Preliminary Lens Model for B1152+199} \label{sec:1152-mod}

B1152+199 is a two component system, and current imaging data provides only
three active constraints on the modeling -- two relative coordinates and a
flux density ratio. Furthermore, since B1152+199 is flat-spectrum, the true
magnification ratio may be masked by variability (though the consistency
between the 4.85 GHz flux densities in GB6 and the VLA data of
Laurent-Muehleisen et al. argues against strong variability).  The modeling of
this system will therefore be postponed until additional constraints can be
obtained, such as the location of the lensing galaxy relative to the images
from optical or infrared observations, or substructure seen in higher
resolution radio maps.

\section{CLASS B1359+154} \label{sec:1359}

CLASS B1359+154 was selected as GB6 J14016+1513, with a flux density
of 66 mJy at 4.85~GHz.  At 1.4~GHz, the NVSS flux density is 115 mJy,
giving an overall low-frequency 
spectral index of $\alpha_{1.4}^{4.85} = -0.447$, near the limit of
our spectral index cutoff ($\alpha_{1.4}^{4.85} = -0.5$).

The X-band discovery 30-second snapshot (not shown) was taken on 16 March
1998. Follow-up 5-minute X-band and 10 minute U-band observations were made on
5 April 1998.  The deep X-band image is shown in Fig.~\ref{fig:1359} (a)
with uniform weighting to enhance the resolution ($0\farcs21$). The map of
the U-band data shown in Fig.~\ref{fig:1359} (b) at a resolution of
$0\farcs15$ was made using natural weighting to show the fainter components.
A total of six components are seen in the deeper X-band image, with an
additional component (F) found near the center component (E).  This component
was later found using model-fitting in the original X-band snapshot, and
appears in the tables.  We tentatively identify the outer components, 
labeled A--D clockwise from the brightest (North) component, as the 
images of a single background source.  The two central components, E and F,
are both found to be extended in the deep X-band dataset, supporting the
hypothesis that these are due to lensing galaxy emission. The positions and
flux densities of the B1359+154 radio components are listed  in
Table~\ref{tab:1359}. Positions were computed without any self-calibration
applied to the data, to minimize biases introduced by those steps. 

The total flux density of the components in the deep X-band image is
$27.9$~mJy, significantly less than the GB6 flux density, possibly indicating
the presence of extended emission that is resolved out of the VLA images.  For
the brightest five components (A--E) detected in both of the 5 April 1998
images, the overall high-frequency spectral index is $\alpha_{8.46}^{14.94} =
-0.9$.  Note that the central component E has a spectral index flatter
than that of the outer components, also indicating that it is probably 
associated with the lensing galaxy, rather than being an additional lensed
image. Component F is too faint to see in the U-band image (it would be less
than $2\sigma$ if it is as flat-spectrum as E).

As in the case of B1152+199, wide-field images were made of the data to
look for companion sources.  In the deep X-band image, for example, no
sources are found above a flux density of $0.5$ mJy within $\pm 128\arcsec$ of
the location of B1359+154.

\subsection{Optical Follow-up of B1359+154} \label{sec:1359-opt}

The spectrum of B1359+154 was extracted from a single 480-second
exposure with the slit at PA $= 70\arcdeg$. This is shown in
Fig.\,\ref{fig:1359spec}.  Several emission lines are observed  with almost no
detected continuum.  These emission lines can all be identified with a source
at redshift $z_s = 3.235 \pm 0.002$.  The lines seen are Ly$\alpha$ at 5150\AA,
\ion{N}{5} $\lambda$ 1240\AA\ at 5252\AA, \ion{C}{4} $\lambda$ 1549\AA\ 
at 6556\AA, \ion{He}{2} $\lambda$ 1640\AA\ at 6948\AA, and \ion{C}{3} 
$\lambda$ 1909\AA\ at 8073\AA.  No features associated with the lensing 
galaxy are observed.

A wide-field ($35'' \times 35''$) Gunn $g$ band COSMIC image of B1359+154 is
shown in Fig.\,\ref{fig:1359img_big}, while a closeup $g$ band image is shown
in Fig.\,\ref{fig:1359img_small}.  The quad structure is clearly observed, and
agrees well with the orientation seen in the radio images.  However, the
seeing and image quality are insufficient to separate out the images and
possible lensing galaxy emission. This will likely have to await Hubble Space
Telescope imaging.  The total magnitude of the system (images plus lens
emission) is $22.5$ in Gunn $g$ and $21.9$ in Gunn $i$.
 
\subsection{Preliminary Lens Model for B1359+154} \label{sec:1359-mod}

The modeling of a gravitational lens system with $N$ unresolved components is
constrained by $2(N-1)$ relative image coordinates and $N-1$
magnification (flux density) ratios. The VLA maps of B1359+154 therefore
provide $9$ modeling constraints. Image positions and flux densities are
taken from the deep X-band data. We assume $q_0 = 0.5$, $H_0 = 100h$
km s$^{-1}$ Mpc$^{-1}$, and a lens redshift of $z_l = 0.5$ for all 
calculations. 

Optimization was performed using a pseudo-image plane minimization technique,
as described in Kochanek 1991\markcite{kochanek91}. Once the model has 
converged, the lens equation is inverted to solve for the true images of
the recovered source position. Coordinates are shifted to fix component A at
(0,0). A $\chi^2$ is calculated using the relative positions
($\theta_i$) and magnification ratios ($r_i = |S(i)/S(A)|$) of the resulting
images, 
$$\chi^2 = \sum_{i=B,C,D} \left[ {(\theta'_i - \theta_i)^2 \over
\Delta \theta_i^2} + {( r_i' - r_i)^2 \over \Delta r_i^2}   \right]$$
where primed quantities are model-predicted and unprimed quantities
are observed. The overall quality of the model is determined by
$\chi^2/{\rm NDF}$, where NDF is the number of degrees of freedom. 

The simplest model consists of a background source being lensed by an
isolated foreground galaxy.  We therefore attempted to model B1359+154 using a
singular isothermal ellipsoid (SIE; Kormann et al. 1994\markcite{kormann}).
An SIE model is described by five parameters (the lens plane coordinates,
velocity dispersion, axial ratio and position angle), so NDF $= 4$. 
Unfortunately, the best possible model using an isolated SIE deflector
proves unsatisfactory due to a remarkably poor positional mismatch.  Models
including external shear of constant magnitude and direction (NDF $= 2$) also
fail to offer a sufficient fit. We thus conclude that a single galaxy, even in
the presence of significant external shear, is unlikely to account for the
observed image configuration of B1359+154.  

A more advanced model may be constructed by adding a second deflector,
parametrized by a singular isothermal sphere (SIS), at the same redshift as
the SIE. Since an SIS is described completely by its velocity dispersion and
position, NDF $= 1$ for a compound SIE+SIS model. (Note that the coordinates
of each deflector are left as free parameters.) Such models provide excellent
fits to the image positions, but fail to reproduce the flux density ratios to
the accuracy required by the deep X-band dataset. Source variability may
however lead to confused estimates of the magnification ratios from a single
epoch of radio data. If we artificially set the flux density ratio
uncertainties at $20\%$ to account for possible variability, an acceptable fit
can be obtained. This procedure is extremely ad hoc, as there is presently no
evidence for variability in B1359+154. Moreover, the rather steep spectral
indices of the lensed components make it unlikely that the source is strongly
variable. We expand the error bars merely to illustrate that B1359+154 can be
modeled as a gravitational lens system. The parameters of a possible SIE+SIS
model ($\chi^2/$NDF$ = 1.4$) are listed in Table~\ref{tab:1359model}.
Model-predicted image positions, magnification factors and time delays are
listed in Table~\ref{tab:1359dev}. Critical curves and caustics for the model
are displayed in Fig.\,\ref{fig:1359crit}. Note that neither the SIE nor SIS
coordinate centers can be identified with the extended radio components E and
F. Fixing the SIE at either location leads to extremely poor models.

If core radii are fit to each of the deflectors, it is possible to reproduce
the flux density ratios to within the true error bars derived from the deep
X-band observation. These models are underconstrained (NDF $= -1$), however,
and we will not discuss them here. 

Additional information is clearly required to construct a viable lens model
for B1359+154. The above SIE+SIS model is merely a first attempt to
describe the system, and will likely be modified as high resolution optical
and radio data are obtained. However, we believe that the above exercise has
demonstrated two important points. First, B1359+154 may be modeled as a
gravitational lens system, though constraints provided by the flux densities
must be relaxed to obtain a sufficient fit. Second, the deflector system for
this lens is almost certainly compound, as single galaxy models are strongly
excluded by our findings. We therefore predict that HST imaging will reveal a
second deflector, making  B1359+154 similar to CLASS B1608+656 (Koopmans \& 
Fassnacht 1999\markcite{koopfass99}). 

\section{Discussion} \label{sec:end}

The source CLASS B1152+199, discovered in the 1998 CLASS-3 observations, is
clearly a lens.  With two bright compact components exhibiting identical flat
radio spectra, it is unlikely that these are two components of the same radio
source, and even more unlikely that this is a chance superposition of two
unassociated radio galaxies.  Furthermore, the optical spectra obtained with
the Keck-II LRIS show unambiguous lines from systems at two distinct
redshifts.  We interpret these as a lensing galaxy at $z_{\ell} = 0.439$ and a
background source (presumably a quasar) at $z_s = 1.019$.

Although only a single redshift component was identified in the LRIS spectrum
of CLASS B1359+154, which we assume is due to the background lensed source at
$z_s = 3.235$, the unusual morphology exhibited in the radio maps almost
certainly points toward a lensing origin.  The geometry, compactness and
identical spectra of the four outer components is compelling evidence for this
hypothesis, while the extended dual central components can be reasonably
identified with emission from the lensing galaxy or galaxies.  The
similarities to the lenses CLASS B1608+656 in configuration (Myers et al.\
1995\markcite{myers95}) and CLASS B2045+265 with lens emission (Fassnacht et
al.\ 1999\markcite{2045}) are striking.  Although the optical image does not
clearly resolve the structure, it is similar enough to the radio geometry to
confirm this system as a gravitational lens.  To test this hypothesis, we
attempted to model B1359+154 as a gravitational lens system.  Though our
results are very preliminary, we believe that a single deflector model can be
excluded by our findings. A compound SIE+SIS deflector provides an adequate
fit if the uncertainties in the flux density ratios are set at $20\%$ to
account for possible variability. We therefore predict that better optical and
IR images of this field will show a second lensing galaxy. Constraints derived
from such images are required to construct a more believable mass model for
this system. 

If it is indeed a lens, the prospects for using CLASS B1152+199 to
measure time delays, and thus the Hubble Constant $H_0$, are good.  It is
bright, well separated, and likely to show measurable variability, having a
flat spectrum ($\alpha_{8.46}^{14.94} = -0.32$).  The prospects for optical
or IR monitoring are poor, due to the lack of a second image in the Palomar 
data.  However, the ROSAT detection of X-ray emission from this source 
suggests the possibility of monitoring by AXAF.

On the other hand, CLASS B1359+154 is weak, and the steep spectral index
($\alpha_{8.46}^{14.94} = -1.00$ for the brightest component) makes it
unlikely that this source will be strongly variable.  However, when combined
with the other lenses previously found in the JVAS and CLASS surveys, both
these new lens systems will add to the statistical evidence placing limits on
the values of cosmological parameters such as the matter density $\Omega_0$ and
cosmological constant $\Lambda_0$ (eg. Falco et al.\ 1998\markcite{falco98}).

\acknowledgments

STM is supported by a Alfred R. Sloan Fellowship.  The CLASS survey at Caltech
is supported by NSF grant AST-9420018 and AST-9117100.  CLASS at Jodrell Bank,
UK, is supported by European Commission, TMR Programme, Research Network
Contract ERBFMRXCT96-0034 ``CERES''.  We thank the staff of the VLA, Palomar
and Keck observatories for their assistance during our observing runs.
This research has made use of the NASA/IPAC Extragalactic Database (NED) which
is operated by the Jet Propulsion Laboratory, California Institute of
Technology, under contract with the National Aeronautics and Space
Administration.  The NASA ADS and ADC were also used during the course
of the work reported here.  We also thank Brian Mason for computing the ROSAT 
X-ray flux for B1152+199.

\clearpage


\begin{table*}
\caption{CLASS B1152+199: Component Positions and Flux
Densities}\label{tab:1152} 
\medskip
{\small
\begin{tabular}{ c c c c c c c }
\hline \hline
Comp & \sc Offset (E) & \sc Offset (N) & $S_X$ (mJy) & $S_U$ (mJy) &
$\alpha_{8.46}^{14.94}$ & $S_X(A)/S_X$ \\ 
\hline \\
A & $0$ & $0$ & $52.27$ & $43.78$ & $-0.31\pm 0.01$ & $1.00$\\
B & $+0\farcs935 \pm 0\farcs005$ & $-1\farcs248 \pm 0\farcs005$ & $17.23$ &
$14.34$ & $-0.32 \pm 0.03$ & $3.03 \pm 0.03$\\
C & $-23\farcs085 \pm 0\farcs036$ & $-0\farcs144 \pm 0\farcs038$ & $2.34$ &
$1.20$ & $-1.2 \pm 0.4$ & $22.3 \pm 1.5$\\[1ex]
\hline
\end{tabular}}
\tablecomments{Positions are offset from RA 11 55 18.2971 and Dec +19 39
42.230 (J2000). The rms noise levels are $0.16$ mJy/beam for the 03 Apr 1998
X-band observation and $0.25$ mJy/beam for the 05 Apr 1998 U-band
observation.}
\end{table*}

\begin{table*}
\caption{CLASS B1359+154: Component Positions and Flux
Densities}\label{tab:1359} 
\medskip
{\small
\begin{tabular}{ c c c c c c c c }
\hline \hline
Comp & \sc Offset (E) & \sc Offset (N) & $S_X$ (mJy) & $S_{Xdeep}$ (mJy) &
$S_U$ (mJy) & $\alpha_{8.46}^{14.94}$ & $S_X(A)/S_X$ \\ 
\hline \\
A & $0$ & $0$ & $9.86$ & $9.57$ & $5.43$ &  $-1.00\pm 0.06$ & $1.00$\\
B & $-0\farcs489 \pm 0\farcs006$ & $-1\farcs256 \pm 0\farcs006$ & $5.87$ &
$5.82$ & $3.51$ &  $-0.89\pm 0.09$ & $1.64 \pm 0.02$\\
C & $-0\farcs310 \pm 0\farcs004$ & $-1\farcs662 \pm 0\farcs004$ & $7.95$ &
$7.92$ & $4.77$ &  $-0.89\pm 0.07$ & $1.21 \pm 0.01$\\
D & $+0\farcs958 \pm 0\farcs019$ & $-1\farcs372 \pm 0\farcs018$ & $2.22$ &
$1.93$ & $1.05$ &  $-1.07\pm 0.30$ & $4.96 \pm 0.17$\\
E & $+0\farcs617 \pm 0\farcs017$ & $-1\farcs138 \pm 0\farcs016$ & $2.49$ &
$2.17$ & $1.49$ &  $-0.66\pm 0.21$ & $4.41 \pm 0.14$\\
F & $+0\farcs437 \pm 0\farcs065$ & $-0\farcs955 \pm 0\farcs064$ & $0.74$ &
$0.50$ & $...$ &  ... & $19.1 \pm 2.5$\\[1ex]
\hline
\end{tabular}}
\tablecomments{Positions are offset from RA 14 01 35.5495 and Dec +15 13
25.643 (J2000). The rms noise levels are $0.224$ mJy/beam for the 16 Mar 1998
X-band observation, $0.066$ mJy/beam for the 05 Apr 1998 deep X-band
observation, and $0.176$ mJy/beam for the 05 Apr 1998 U-band observation. 
The deep X-band data is used in the calculation of flux density ratios and
spectral indices. Using the 16 Mar 1998 X-band data yields similar results.}
\end{table*}

\clearpage

\begin{table*}
\caption{CLASS B1359+154: SIE+SIS Model Parameters} \label{tab:1359model}
\medskip
{\small
\begin{tabular}{ c c c c c c }
\hline \hline
  Comp & \sc Offset (E) & \sc Offset (N) & $\sigma$ (km/s) & PA (deg) &
$b/a$ \\ \hline \\ 
 SIE & $+0\farcs3618$ & $-0\farcs9773$ & $210.13$ & $135.29$ & $0.694$ \\
 SIS  & $-1\farcs5255$ & $-1\farcs4969$ & $137.63$ \\
 Source & $-0\farcs0544$ & $-1\farcs0037$\\
\hline
\end{tabular}}
\end{table*}

\begin{table*}
\caption{CLASS B1359+154: Model-predicted image positions, magnification
factors and time delays} \label{tab:1359dev}
\medskip

{\small
\begin{tabular}{ c c c c c }
\hline \hline
 Comp & \sc Offset (E) & \sc Offset (N) & $\mu$ & $\Delta t$ ($h^{-1}$ days) \\
\hline \\
 A & $+0.0000$ & $+0.0000$ & $+11.30$ & $0.0$\\  
 B & $-0.4887$ & $-1.2560$ & $-7.60$ & $2.7$\\  
 C & $-0.3100$ & $-1.6622$ & $+7.44$ & $2.5$\\  
 D & $+0.9579$ & $-1.3721$ & $-2.40$ & $10.4$\\[1ex]
 
\hline
\end{tabular}}
\end{table*}

\clearpage


\clearpage

\figcaption{The gravitational lens CLASS B1152+199. (a) 
8.46 GHz VLA radio image, $0\farcs21$ resolution, uniform weighting. 
(b) 14.94 GHz VLA image, $0\farcs12$ resolution, uniform 
weighting.\label{fig:1152} }

\figcaption{Keck-II LRIS spectrum of B1152+199 dominated by light from the 
background source. The strong broad lines establish the source redshift as
$z_s = 1.019$.  The absorption lines at the blue end are associated with 
the lens.\label{fig:1152spec1} }

\figcaption{Keck-II LRIS spectrum of lensing galaxy in B1152+199.
The lines from the lens give $z_{\ell} = 0.439$.  Lines associated 
with the background source are also seen in this spectrum.
\label{fig:1152spec2} }

\figcaption{The COSMIC $g$ band contour map of B1152+199, showing the 16.5-mag
stellar object at the lens position.  Contours are at $10\%$ intervals, and
there is no sign of a secondary image.  
The field-of-view displayed is $5'' \times 5''$. 
\label{fig:1152img} }

\figcaption{The gravitational lens CLASS B1359+154. (a) 
Deep 8.46 GHz VLA radio image, $0\farcs21$ resolution, uniform weighting.
(b) 14.94 GHz VLA image, $0\farcs15$ resolution, natural weighting. 
\label{fig:1359} }

\figcaption{Keck-II LRIS spectrum of B1359+154.
The identified emission lines give a source redshift of
$z_s = 3.235$.\label{fig:1359spec} }

\figcaption{Wide field ($35'' \times 35''$) COSMIC $g$ band image of
CLASS B1359+154.  The object is centered in the frame.
\label{fig:1359img_big} }

\figcaption{Close-up COSMIC $g$ band image of CLASS B1359+154.
The pixel size is $0\farcs286$.
\label{fig:1359img_small} }

\figcaption{Critical curves (solid lines), caustics (inset), and curves of
constant time delay (dashed lines) for the best-fit SIE+SIS model of CLASS
B1359+154. 
\label{fig:1359crit} }

 
\clearpage

\begin{figure}
\epsfxsize=3in
\epsfbox[33 55 582 740]{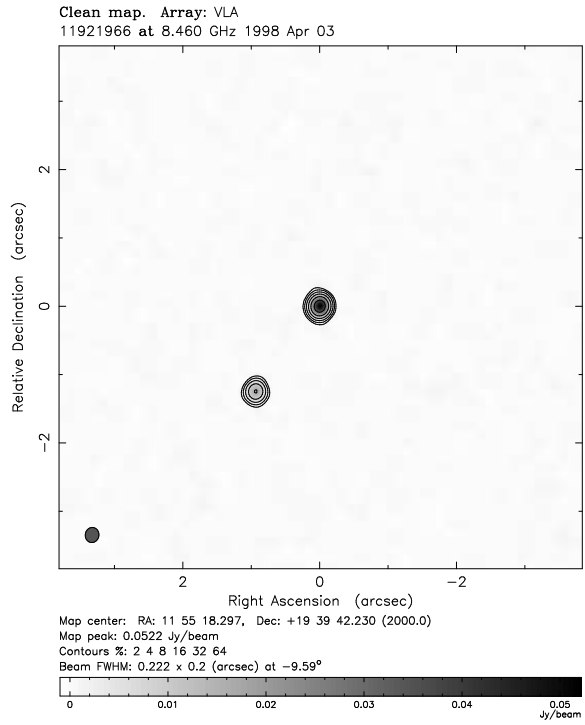}
\figurenum{1}
\caption{(a) 8.46 GHz VLA image of CLASS B1152+199.}
\end{figure}

\begin{figure}
\epsfxsize=3in
\epsfbox[33 55 582 740]{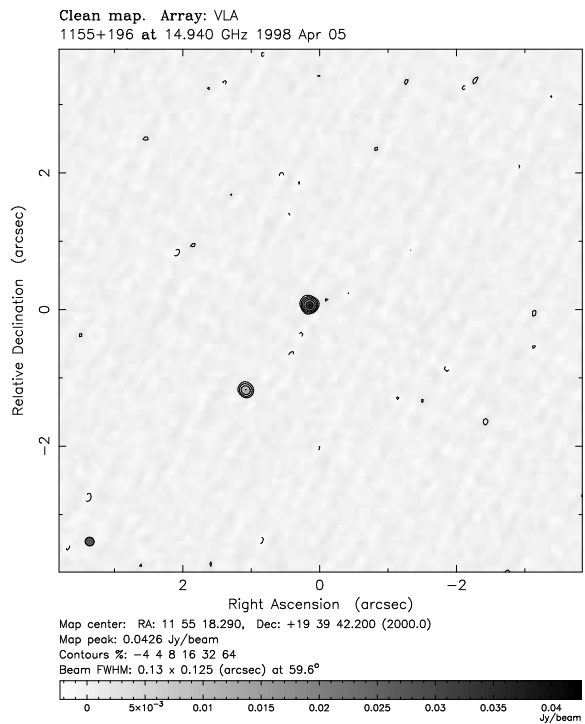}
\figurenum{1}
\caption{(b) 14.94 GHz VLA image of CLASS B1152+199.}
\end{figure}

\begin{figure}
\epsfxsize=3in
\epsfbox[39 24 536 350]{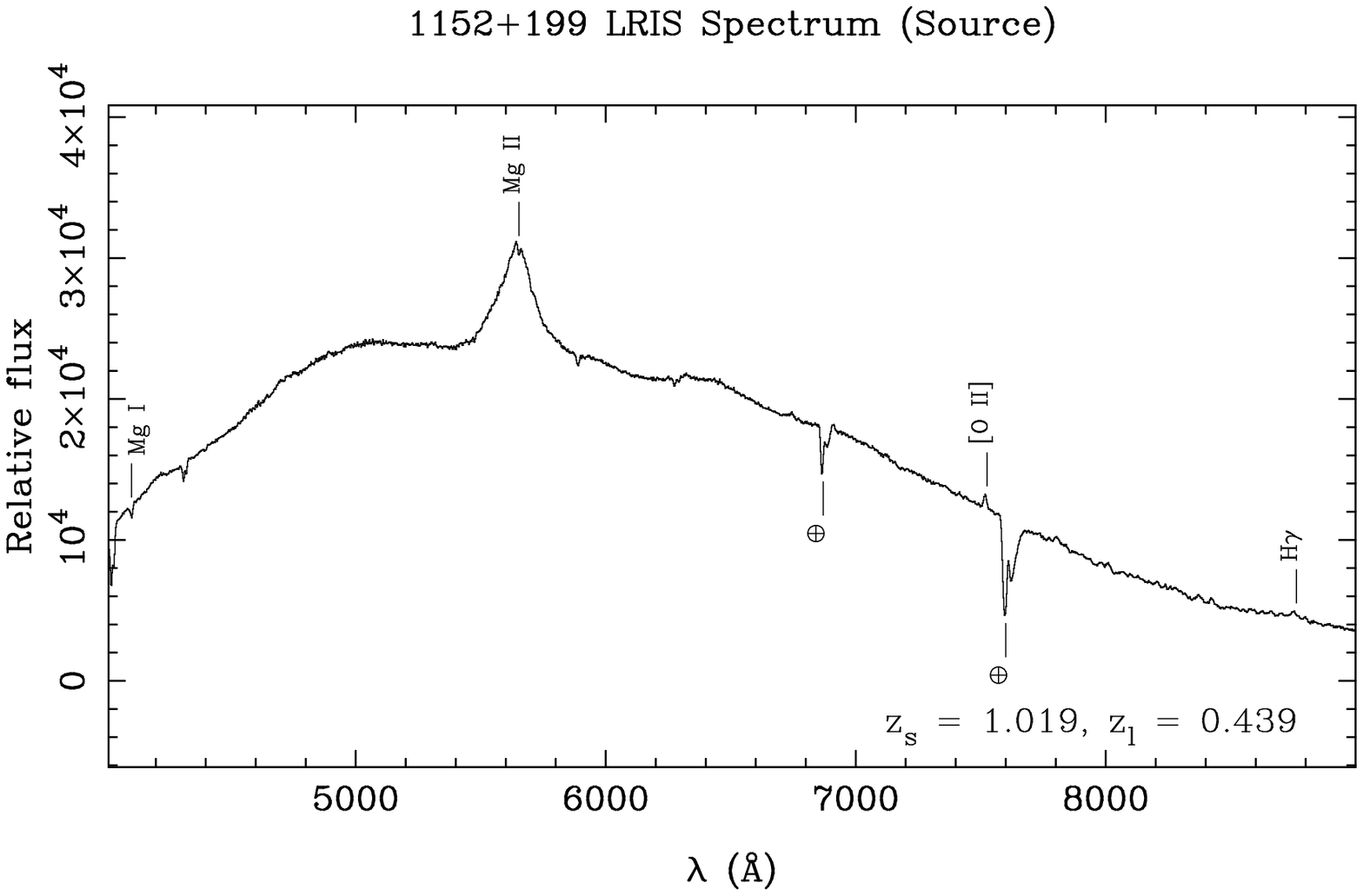}
\figurenum{2}
\caption{Keck-II LRIS spectrum of background source in B1152+199.}
\end{figure}

\begin{figure}
\epsfxsize=3in
\epsfbox[39 24 550 350]{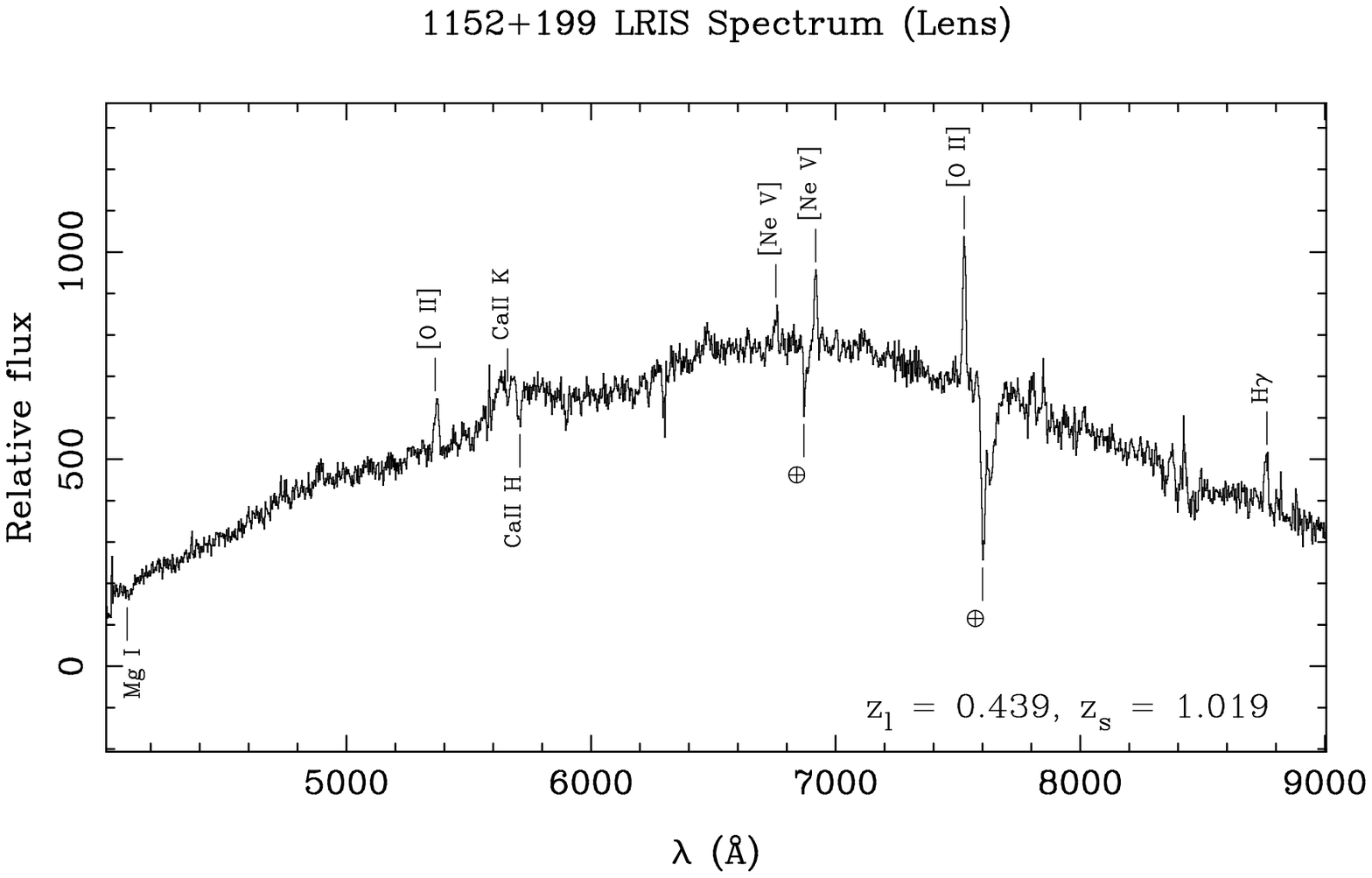}
\figurenum{3}
\caption{Keck-II LRIS spectrum of lensing galaxy in B1152+199.}
\end{figure}

\begin{figure}
\epsfxsize=3in
\epsfbox{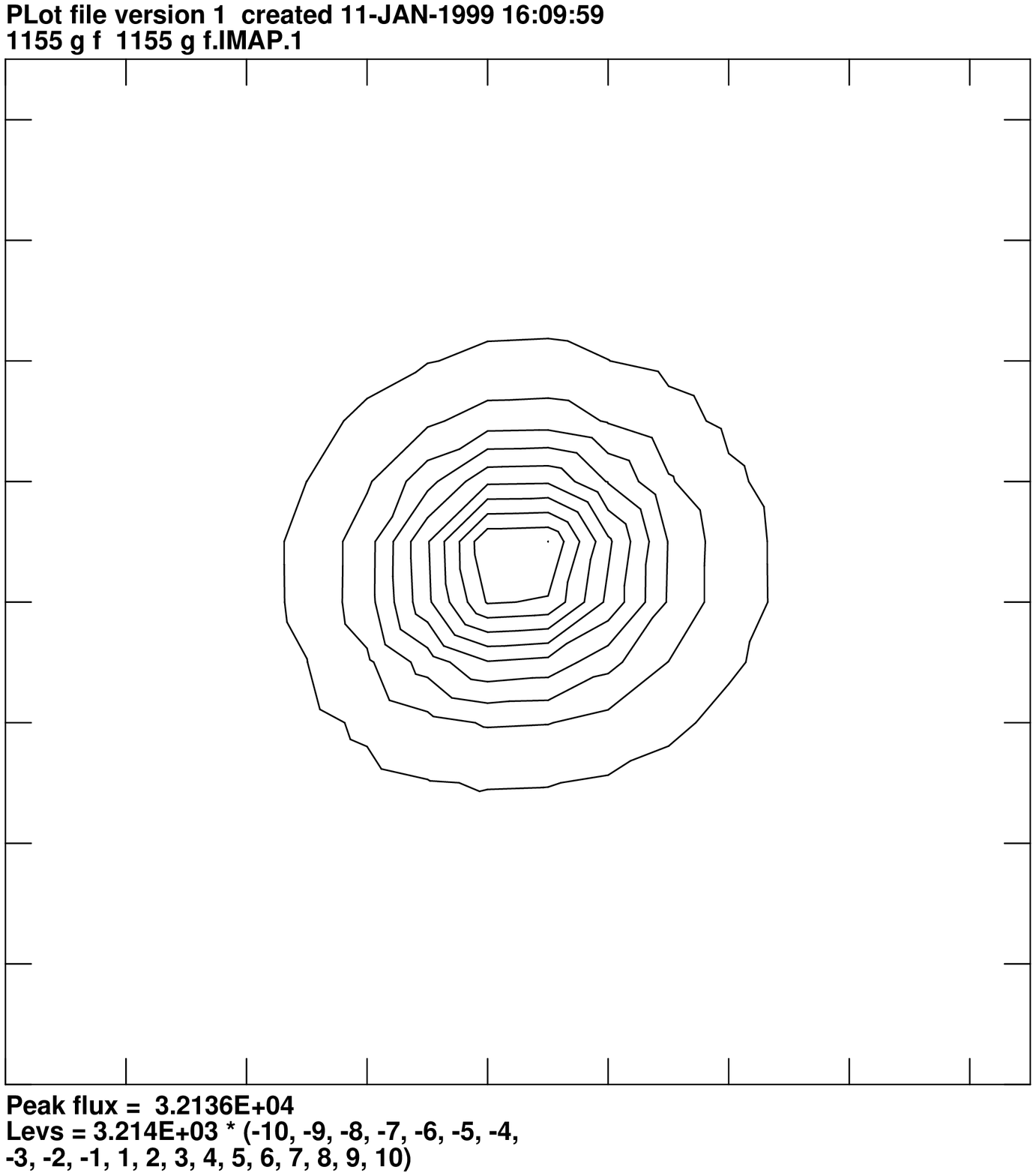}
\figurenum{4}
\caption{Palomar Gunn-$g$ contour map of B1152+199.}
\end{figure}

\clearpage

\begin{figure}
\epsfxsize=3in
\epsfbox[33 55 582 740]{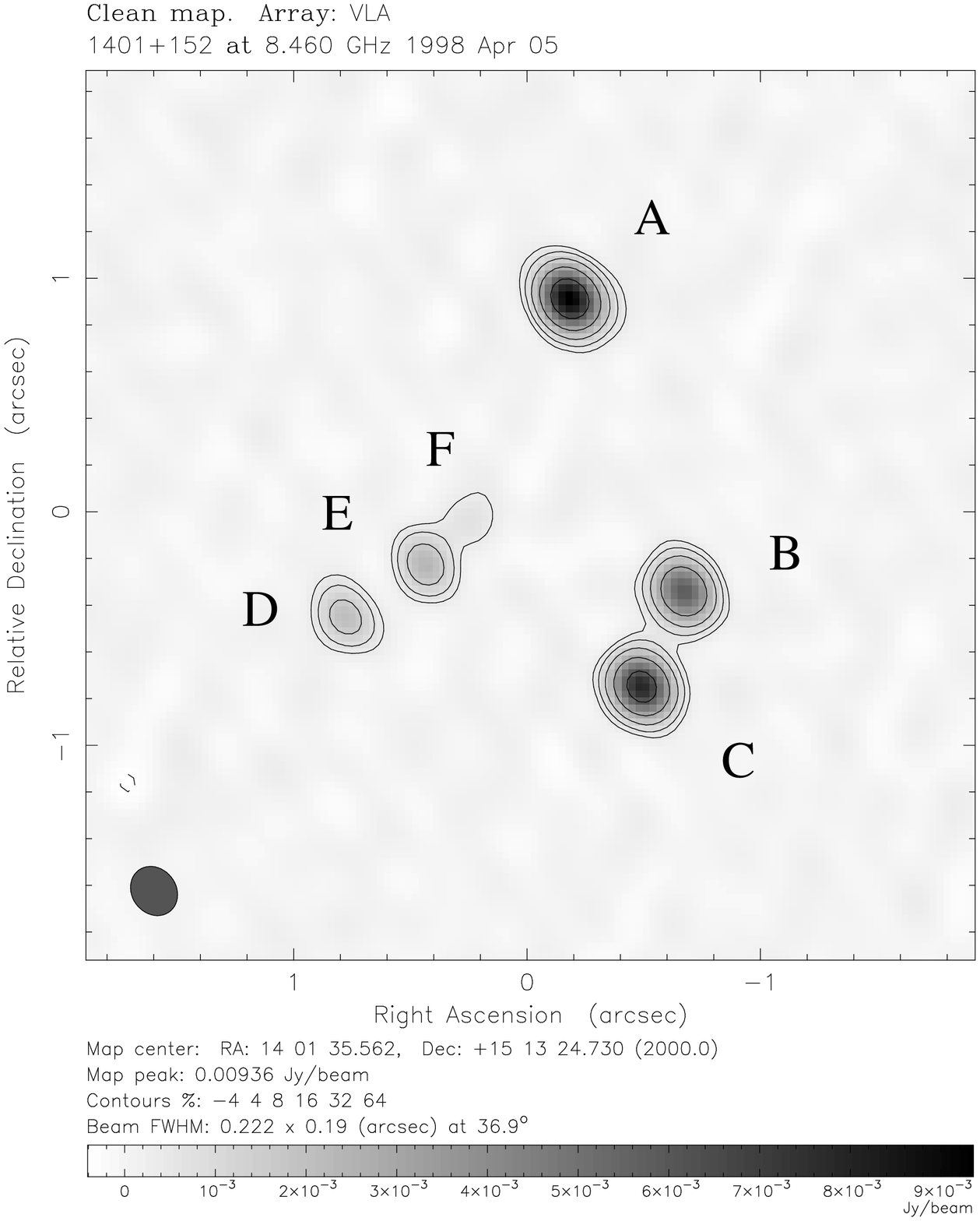}
\figurenum{5}
\caption{(a) Deep 8.46 GHz VLA image of CLASS B1359+154.}
\end{figure}

\begin{figure}
\epsfxsize=3in
\epsfbox[33 55 582 740]{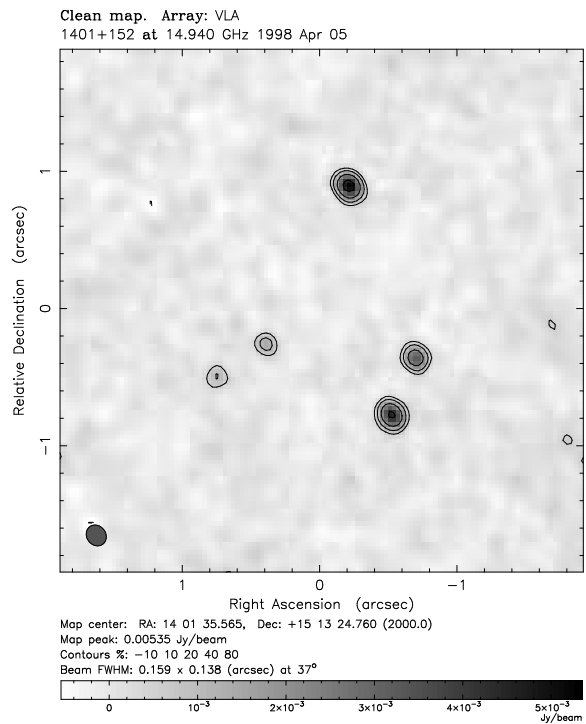}
\figurenum{5}
\caption{(b) 14.94 GHz VLA image of CLASS B1359+154.}
\end{figure}

\begin{figure}
\epsfxsize=3in
\epsfbox[39 24 550 350]{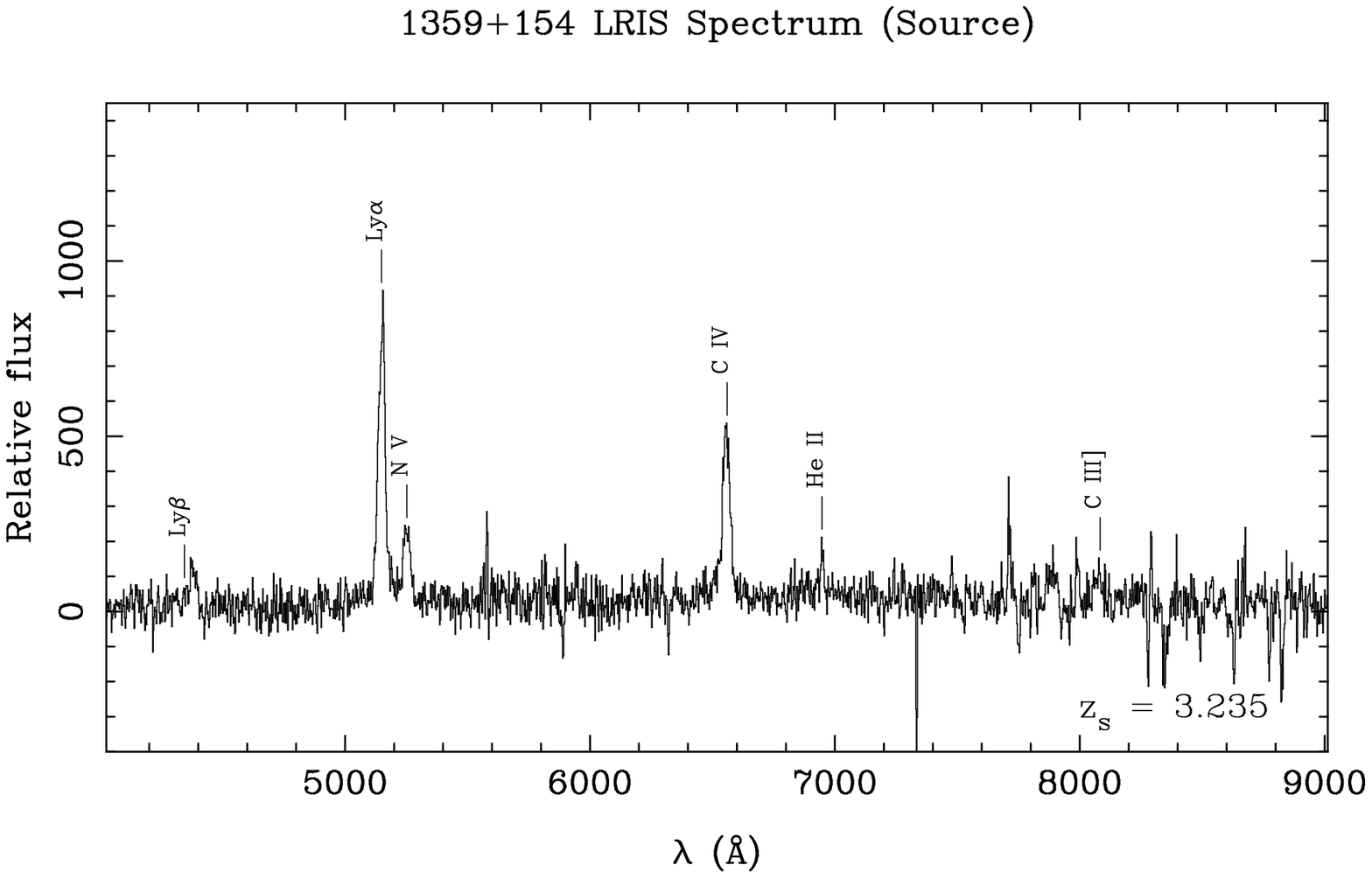}
\figurenum{6}
\caption{Keck-II LRIS spectrum of B1359+154.}
\end{figure}

\begin{figure}
\epsfxsize=3in
\epsfbox{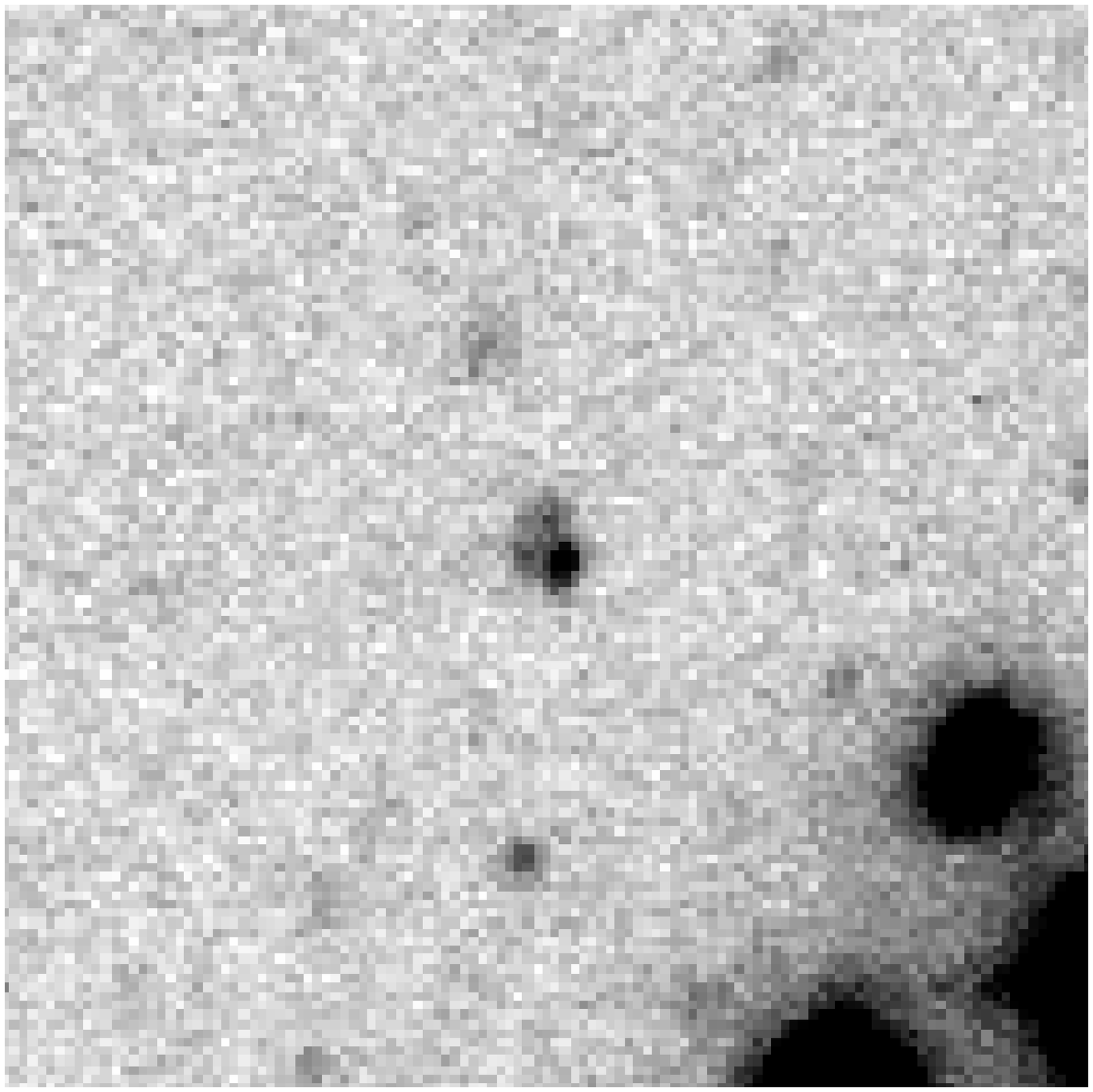}
\figurenum{7}
\caption{Wide field Palomar Gunn-$g$ image of B1359+154.}
\end{figure}
 
\begin{figure}
\epsfxsize=3in
\epsfbox{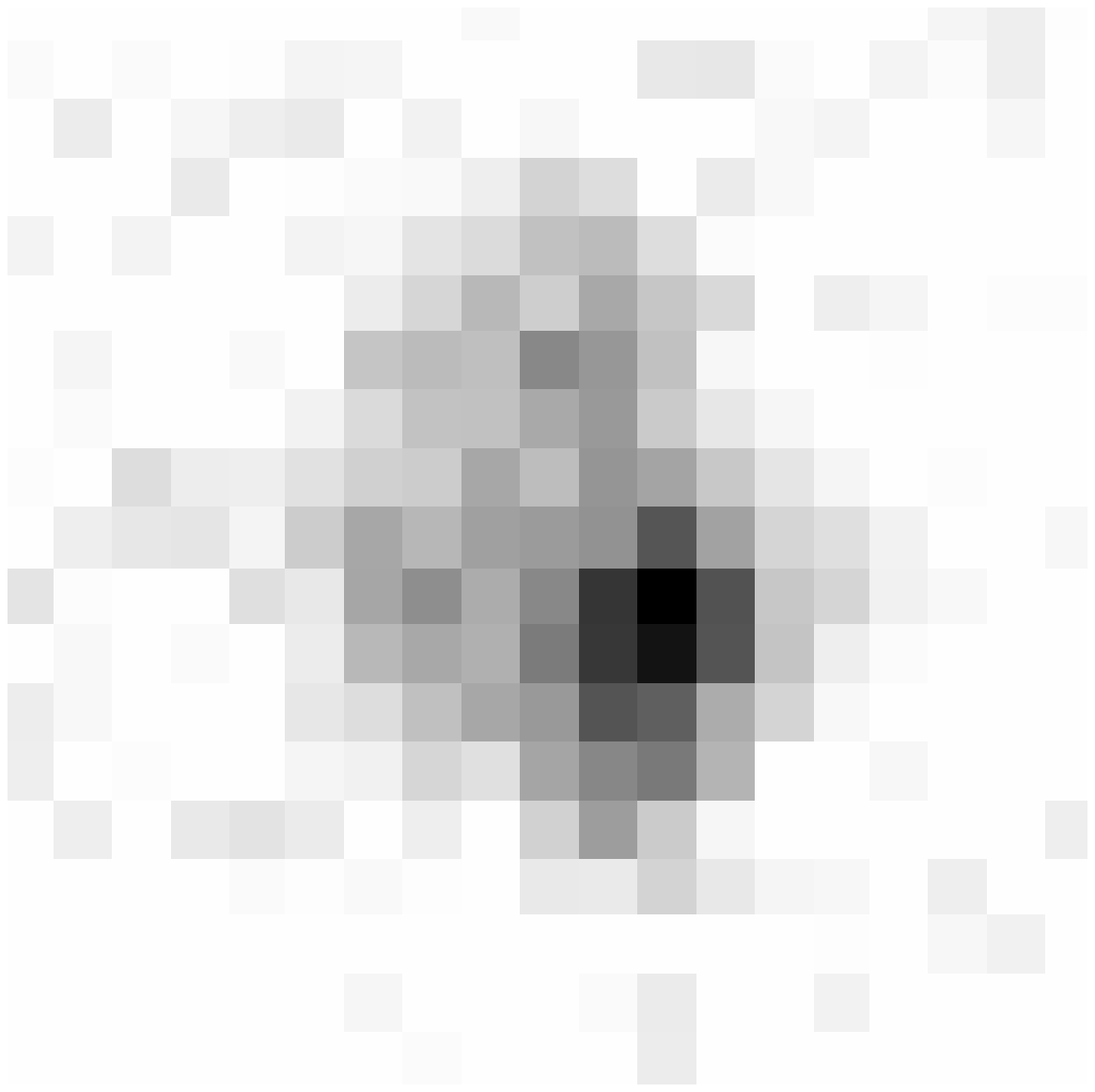}
\figurenum{8}
\caption{Palomar Gunn-$g$ image of B1359+154.}
\end{figure}

\begin{figure}
\epsfxsize=3in
\epsfbox[48 33 545 530]{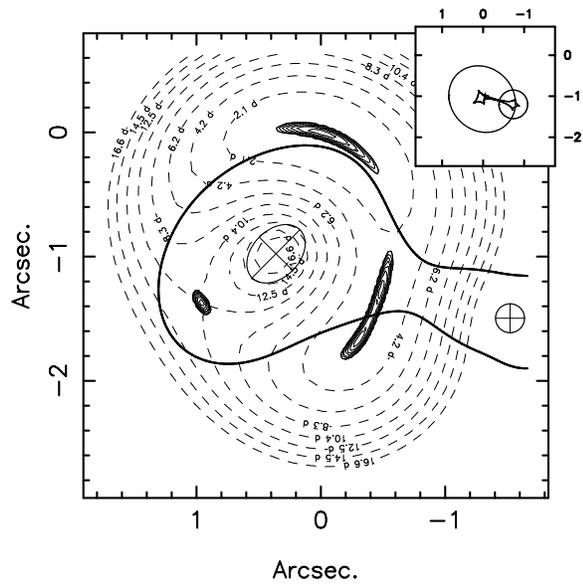}
\figurenum{9}
\caption{Critical lines and caustics for the best-fit SIE+SIS model of
B1359+154.}
\end{figure}

\end{document}